# Intracavity Dual-Resonance Stimulated Raman Spectroscopy with Cavity Ringdown Readout for Resolving Hydrogen Rotational Raman Transitions

Qinxue Nie, Guanda Lyu, Yue Yan, and Wei Ren*

*Department of Mechanical and Automation Engineering, The Chinese University of Hong Kong, New Territories, 999077, Hong Kong SAR, China*

**Corresponding author: renwei@mae.cuhk.edu.hk*

Gas-phase rotational Raman lineshape metrology of hydrogen ($H_2$) is challenging due to its weak Raman scattering cross-sections and intrinsically narrow linewidths. We report the first measurement of the complete Dicke-narrowing evolution of the $H_2$ rotational Raman transitions $S_0(1)$ and $S_0(0)$, enabled by intracavity dual-resonance stimulated Raman spectroscopy with cavity ringdown readout and kHz-level spectral resolution. Our results establish a benchmark for gas-phase Raman lineshape measurements and provide stringent constraints for regime-based pressure-dependent linewidth models.

Raman scattering from infrared-inactive molecules such as hydrogen ($H_2$) has been studied for decades, enabling stringent tests of molecular theory and fundamental constants [1–4] and supporting applications in astrophysics [5,6] and in the energy sector [7–10]. Despite substantial progress in spontaneous and stimulated Raman measurements, no existing spectroscopic tool simultaneously provides the spectral resolution and sensitivity required to clearly resolve the intrinsically narrow Raman lineshapes of $H_2$ at low pressures, particularly across the full Dicke-narrowed regime [3,11–14].

Dicke narrowing emerges when the molecular mean free path approaches the characteristic length scale set by the radiation field, such that velocity-changing collisions partially suppress Doppler broadening [15]. Measuring the resulting lineshape evolution is essential for establishing quantitative, unified descriptions of Dicke narrowing in $H_2$ Raman transitions [3,13], for providing benchmark spectroscopic data for precision $H_2$ molecular physics [16,17], and for interpreting $H_2$ Raman spectra in astrophysical environments [5,6]. Capturing this evolution from the Doppler-limited to the collision-broadened limits therefore requires a Raman spectrometer that combines MHz-level spectral resolution and high detection sensitivity with reliable operation over a wide pressure range, extending into the low-pressure regime.

Tunable single-mode lasers make stimulated Raman scattering (SRS) an attractive method for high-resolution spectroscopy since the Raman resonance can be scanned by tuning either the pump or the Stokes laser wavelength. However, SRS implementations previously applied to Dicke narrowing in $H_2$ generally exhibit instrumental linewidths at the hundreds-of-MHz level [12,13], set by the laser linewidth or the spectral resolution of the Stokes-light detection chain. This resolution is inadequate for directly resolving the rotational S-branch Raman transitions in the narrowing region, where intrinsic linewidths can fall in the tens of MHz. The situation is further complicated by the fact that the S-branch Dicke-narrowed regime occurs at sub-atmospheric pressures, where the SRS gain becomes small and hard to measure [18]. Hence, many studies have been limited to the vibrational $Q_1(1)$ transition with a much larger linewidth (hundreds of MHz) and Raman gain [3,13,18–20]. However, the critical low-pressure interval that links the Dicke-narrowing region to the Doppler-dominated limit has remained out of reach experimentally.

Here we introduce an intracavity dual-resonance stimulated Raman spectroscopy platform that overcomes these limitations by simultaneously enhancing the pump field and the effective Stokes interaction within a high-finesse optical cavity. As shown in Fig. 1(a), the optical cavity resonantly builds up the circulating pump light while extending the effective path length for the Stokes light. This dual enhancement yields ultrahigh detection sensitivity together with a large dynamic range, enabling continuous measurements from the weak, low-pressure Doppler-dominated regime to the high-pressure collision-broadened regime.

To acquire high-resolution Raman lineshapes, we implement difference-frequency scanning by tuning the cavity length while maintaining pump-cavity locking at each cavity length, as shown in Fig. 1(b). This approach decouples the scan resolution from the laser linewidth and can achieve a minimum attainable difference-frequency resolution at the kHz level [21]. Using this approach, we directly measure, for the first time, the complete Dicke-narrowing evolution of the $H_2$ rotational Raman transitions $S_0(0)$ and $S_0(1)$, spanning the Doppler-dominated, Dicke-narrowed, and collision-broadened limits.

The diagnostic method is based on SRS, in which pump light is coherently converted into Stokes light. For a uniform integration region, the fractional Stokes intensity gain is given by [22]:

$$\frac{\Delta I_s}{I_s^0} = \exp\left(g_R I_p L\right) \qquad (1)$$

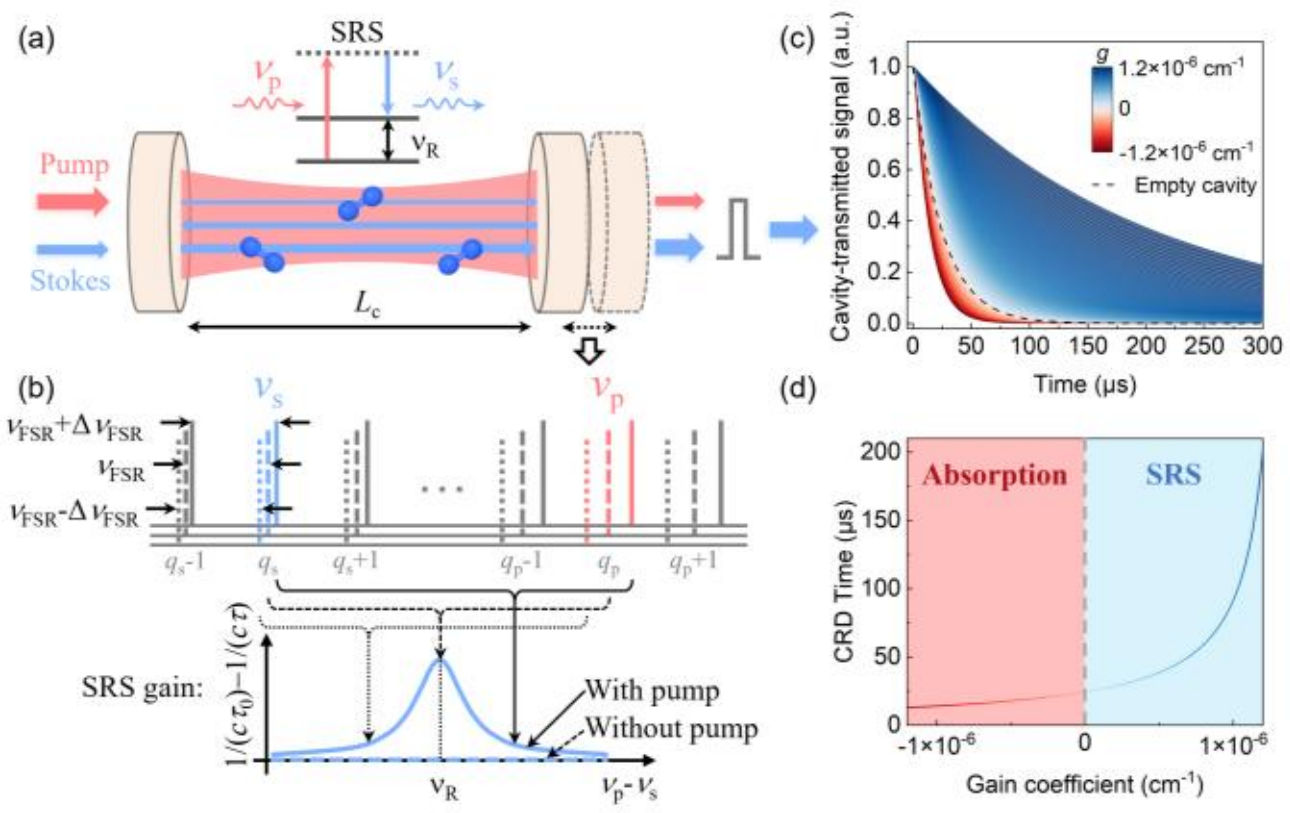

FIG 1. Principle of intracavity dual-resonance stimulated Raman spectroscopy. (a) The pump and Stokes fields are simultaneously resonant with a high-finesse cavity to enhance the SRS interaction. (b) Difference-frequency scanning for high-resolution Raman spectroscopy: the pump ($\nu_p$) and Stokes ($\nu_s$) fields resonate with the cavity longitudinal modes $q_p$ and $q_s$, respectively, and the optical frequency difference is swept across the Raman transition $\nu_R$ by stepping the cavity free spectral range $\nu_{FSR}$ in increments of $\Delta\nu_{FSR}$. (c) The induced Stokes gain is quantified from the Stokes CRD events, which are simulated for different intracavity gain coefficients $g$; negative $g$ denotes net intracavity loss (e.g., caused by molecular absorption). (d) Variation of the extracted ringdown time with $g$.

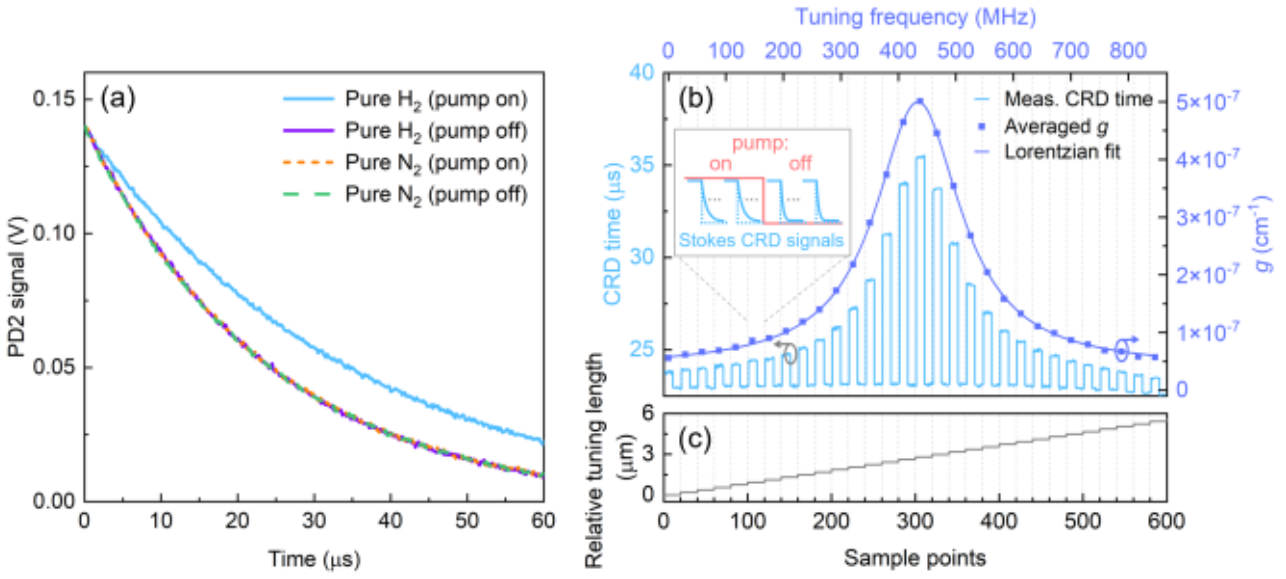

FIG 2. Measurement of Stokes ringdown signal and retrieval of the $H_2$ Raman spectrum. (a) Representative Stokes ringdown transients recorded with the pump laser switched on and off, illustrating the absence of gain in $N_2$ or pump off. (b) Measured $H_2$ $S_0(1)$ spectrum at 298 K and 166.8 kPa, obtained using the pump on-and-off scheme shown in the inset at each cavity-length setting. (c) Stepwise cavity length (or FSR) tuning used to sweep the pump-Stokes difference frequency across the Raman transition.

where $I_S^0$ and $\Delta I_S$ are the initial and incremental Stokes intensities, respectively; $g_R$ is the Raman gain coefficient (proportional to the concentration of the target species); $I_p$ is the pump intensity; and $L$ is the effective pump-Stokes-molecule interaction length. Here, we define the Stokes gain coefficient as $g \equiv g_R I_p$. In our implementation, the optical cavity enhances $I_p$ by more than four orders of magnitude, thereby directly boosting the SRS gain.

We quantify $g$ by monitoring the cavity ringdown (CRD) time of the Stokes light. In conventional CRD absorption spectroscopy, molecular absorption increases the intracavity loss and thus shortens the ringdown time. In contrast, intracavity SRS generates gain at the Stokes frequency, partially compensating the net loss and increasing the Stokes ringdown time. The opposite responses are schematically illustrated in Fig. 1(c,d). The Stokes gain coefficient $g$ can be retrieved from:

$$g = \frac{1}{c}\left(\frac{1}{\tau_0} - \frac{1}{\tau}\right) \tag{2}$$

where $c$ is the speed of light, $\tau_0$ is the ringdown time for the empty cavity, and $\tau$ is the ringdown time in the presence of SRS gain. As depicted in Fig. 1 (d), this gain-CRD scheme transduces the Raman interaction into an increase in ringdown time, providing a favorable measurement signal.

High-resolution Raman spectra are acquired by scanning the pump-Stokes frequency difference using a cavity-based difference-frequency tuning scheme, as shown in Fig. 1(b). Specifically, we finely tune the cavity free spectral range (FSR) to sweep the relative detuning between the pump and Stokes lasers while maintaining resonance. This approach is particularly advantageous for narrow Raman features as the spectral resolution is no longer directly limited by the laser linewidth. Instead, the Raman spectral resolution depends on the frequency difference resolution between the intracavity pump and Stokes lasers.

The spectroscopic method involves the use of two continuous-wave (CW) pump and Stokes lasers, which are effectively coupled into a high-finesse optical cavity for dual-wavelength resonance. The detailed experimental setup is provided in Supplementary Note 1. The wide tuning range (1600 - 1750 nm) of the Stokes laser enables the measurement of two rotational Raman transitions of $H_2$, $S_0(1)$ = 587 cm$^{-1}$ and $S_0(0)$ = 354.36 cm$^{-1}$, while fixing the pump wavelength at 1530.80 nm. The cavity finesse at the pump wavelength is determined to be 146,105 using CRD measurement (Supplementary Note 2). With the maximum incident pump power of 90 mW and the achieved laser-to-cavity coupling efficiency of ~50%, the intracavity pump light intensity can reach up to $5.92 \times 10^9$ W·m$^{-2}$. Similarly, the cavity finesse at the Stokes wavelengths is determined to be 192,164 (1681.94 nm) and 229,879 (1618.60 nm), respectively, corresponding to a Stokes-gas interaction length exceeding 10 km.

The two-photon interaction feature of SRS provides an inherent route to in situ background subtraction. As shown in Fig. 2(a), when the pump-Stokes frequency difference is set to the $S_0(1)$ Raman transition, the Stokes ringdown time increases only when both optical fields and $H_2$ are present in the optical cavity, i.e., when SRS gain is generated. In the absence of the pump field, $H_2$ is effectively transparent to the

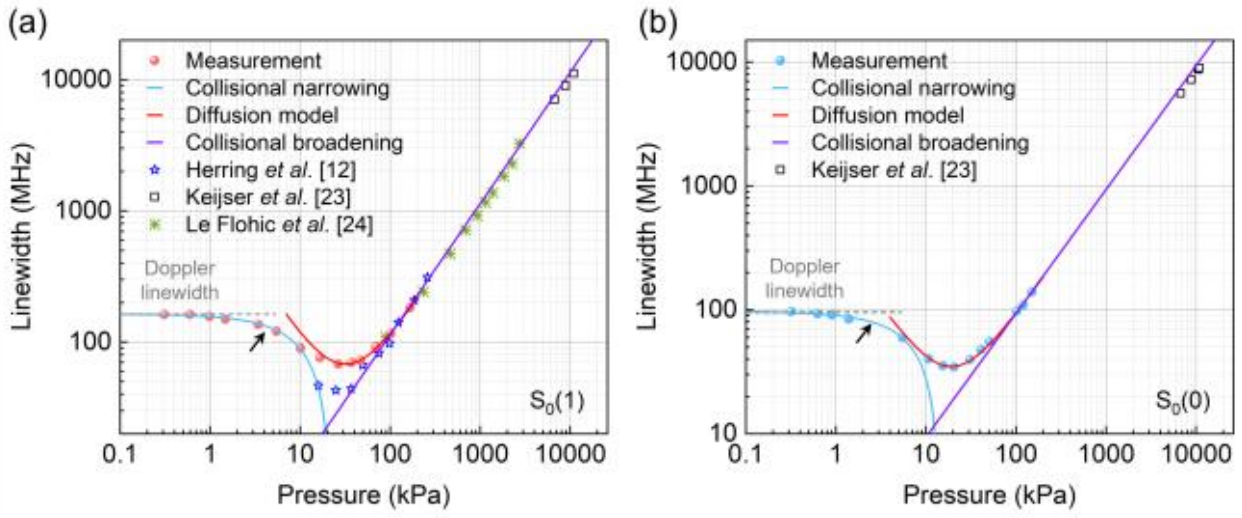

FIG 3. Pressure-dependent linewidths of $H_2$ rotational Raman transitions measured across the full Dicke narrowed regime in pure $H_2$. (a) Variation of FWHM of the $S_0(1)$ transition with pressure. (b) Variation of FWHM of the $S_0(0)$ transition with pressure. Spectra are fitted using three regime-specific models, including collisional narrowing model (Eqn. 3), diffusion model (Eqn. 4) and collisional broadening model (Eqn. 5). Previously reported $S_0(1)$ results [12,23,24] and $S_0(0)$ results [12] are included for comparison.

Stokes light, allowing the direct measurement of the background ringdown time ($\tau_0$) with $H_2$ still in place.

The resolved Raman spectrum is acquired by tuning the pump-Stokes difference frequency via stepwise adjustment of the cavity FSR. This is implemented by scanning the cavity length with a piezoelectric actuator (PZT) mounted on one of the cavity mirrors. Figure 2(b) shows a representative $S_0(1)$ spectrum measured in pure $H_2$ at 298 K and 166.8 kPa with 0.27 mW incident pump power. Note that the low pump power was selected to avoid the self-generated Stokes emission driven by the pump alone, which would obscure the stimulated Raman signal. At each frequency step, real-time background subtraction is realized using the pump on-and-off sequence shown in the inset of Fig. 2(b). The spectrum is then obtained from a stepwise cavity-length scan as illustrated in Fig. 2(c). The step size is adapted to the linewidth of the $H_2$ Raman transition under each experimental condition to ensure adequate sampling of the lineshape. According to the cavity FSR ($\nu_{FSR}$ = 1.328 GHz) and finesse, the full width at half-maximum (FWHM) of the cavity mode is determined to be 9.1 kHz at 1530.80 nm, 5.8 kHz at 1681.94 nm and 6.9 kHz at 1618.60 nm. These linewidths set the minimum attainable Raman spectral resolution, corresponding to 0.49 kHz for $S_0(1)$ and 0.82 kHz for $S_0(0)$ (Supplementary Note 3).

We first studied the complete Dicke-narrowing evolution of the $H_2$ $S_0(0)$ and $S_0(1)$ transitions by filling the cavity with pure $H_2$ at 298 K and varying the pressure from 0.3 kPa to ~170 kPa. The measured spectrum at each pressure is fitted with a lineshape function selected to match the dominant broadening mechanism: a Gaussian profile in the Doppler-dominated limit, a Voigt profile in the intermediate regime where collisional effects become appreciable while Doppler broadening remains significant, and a Lorentzian profile in the collision-dominated regime. In our measurements, Dicke narrowing occurs in the collision-affected regime, and spectra can be fitted with Lorentzian-dominated profiles whose linewidth varies non-monotonically with pressure.

The pressure dependence of the linewidth evolution can be described by three empirical models that are applied in different pressure intervals [12,20,22,25]. At low pressures, the collisional narrowing model accounts for the initial reduction of the effective Doppler width as velocity-changing collisions tend to suppress Doppler broadening. At intermediate pressures, a diffusion model captures the crossover region in which narrowing weakens and collisional broadening becomes increasingly prominent. At high pressures, the linewidth follows a collisional broadening model, reflecting the collision-dominated limit where the width increases approximately linearly with pressure. The three scenarios can be expressed by the following equations:

Collisional narrowing: $$\Delta\nu_R = \nu_D + \gamma_1 P \quad (3)$$

Diffusion model: $$\Delta\nu_R = \frac{A}{P} + BP \quad (4)$$

Collisional broadening: $$\Delta\nu_R = \gamma_2 P \quad (5)$$

where $\Delta\nu_R$ is the linewidth (FWHM) of the Raman transition, $\nu_D$ is the Doppler linewidth, $\gamma_1$ is the collisional narrowing coefficient, $P$ is the gas pressure, and $A$ and $B$ are the self-diffusion coefficient and the density-broadening coefficient, respectively; and $\gamma_2$ is the collisional-broadening coefficient.

TABLE I. Broadening coefficients with fitting uncertainties for $H_2$ $S_0(0)$ and $S_0(1)$ Raman transitions at 298 K.

| | | $\nu_D$ (MHz) | $\gamma_1$ (MHz/atm) | A (MHz·atm) | B (MHz/atm) | $\gamma_2$ (MHz/atm) |
|---|---|---|---|---|---|---|
| *S*0(1) | **This work** | 163.46±1.28 | -756.09±28.31 | 10.75±0.40 | 107.18±1.03 | 114.88±4.97 |
| | Ref. | Theoretical: 154.81 | Not stated | 6.15 [12] | 114±5 [12] | 104±2 [23]<br>100.2±2 [24] |
| *S*0(0) | **This work** | 97.09±1.26 | -698.04±49.15 | 3.30±0.16 | 92.40±0.92 | 94.88±2.37 |
| | Ref. | Theoretical: 93.46 | Not stated | 1.87 [12] | 77±2 [12] | 84±2 [23]<br>82±4 [24] |

Note: The temperatures used in the referenced studies are 295 K (Ref. [12]), 293 K (Ref. [23]), and 302.8 K (Ref. [24]). References [23] and [24] report collisional-broadening coefficients for the $S_0(0)$ transition but do not provide the corresponding linewidths at the measured pressures.

The measured linewidths of $S_0(0)$ and $S_0(1)$ at different pressures are depicted in Fig. 3(a) and (b), respectively, which are overall well fitted by the three empirical models in their corresponding pressure regions. The fit parameters with the corresponding fitting uncertainties are summarized in Table I; representative Raman spectra together with their fits are provided in Supplementary Note 4. However, the diffusion model (Eqn. (4)) does not capture the low-pressure approach to the Doppler-broadened limit. Its applicability is restricted to pressures above a cutoff, which is defined as the point where the linewidth predicted by the diffusion model differs by more than 10% from the prediction of hard-collision lineshape theory [13]. To date, collisional narrowing is more appropriately described by lineshape theories based on hard- and soft-collision models, while $H_2$ collisions are expected to fall between the two extremes [25,26]. Experimentally, the only measured data based on SRS available in the literature are the study of the $S_0(1)$ transition at pressures of 16-260 kPa by Herring *et al.* [12], which are plotted in Fig. 3 (a) for comparison. Although good agreement can be found between the present study and the previous work at pressures above 50 kPa, significant deviations are found at pressures below 50 kPa, with the measured linewidths differing by a factor of about 1.6. It should be noted that the study conducted in Ref. [12] did not measure the linewidth directly from the lineshape profile. Instead, the Lorentzian components were extracted from the Voigt-profile fit, where the Gaussian width of ~100 MHz was dominated by the instrumental linewidth.

We also compared the Doppler linewidths of the two $H_2$ Raman transitions with the theoretical values, as no previous experimental data were available. The theoretical Doppler linewidth is given by [13]:

$$\nu_D = \frac{k_e}{\pi}(\frac{2ln2kT}{m})^{1/2} \quad (6)$$

where $k$ is the Boltzmann constant, $T$ is the temperature, $m$ is the molecular mass; and $k_e$ is the effective SRS wave vector, which is given by:

$$k_e = 2\pi(2\nu_p\nu_s(1-cos\theta)+\nu_R^2)^{1/2} \quad (7)$$

where $\nu_R$ is the Raman shift, $\nu_p$ and $\nu_s$ are the pump and Stokes frequencies, respectively; and $\theta$ is the pump-Stokes crossing angle. Our measured Doppler linewidths are in good agreement with the theoretical values, only differing by 5.6% for $S_0(1)$ and 3.9% for $S_0(0)$ [12]. We attribute this small discrepancy primarily to cavity drift during stepwise scanning and expect that it can be reduced in future measurements by stabilizing the cavity FSR, e.g., by referencing it to a frequency comb [27–29].

Dicke narrowing becomes prominent when the molecular mean free path is comparable to the characteristic spatial scale of the Raman interaction, typically expressed by the criterion $l \sim 1/k_e$ [15]. Applying this criterion yields onset pressures of 4.06 kPa for $S_0(1)$ and 2.45 kPa for $S_0(0)$, respectively, marked by arrows in Fig. 3(a, b). Benefiting from the combined high spectral resolution, high sensitivity and wide dynamic range, we experimentally determine the collisional narrowing coefficient $\gamma_1$ of $H_2$ rotational Raman transitions; to the best of our knowledge, this parameter has not been reported previously from direct linewidth measurements. The linewidth reaches minima of 68 MHz for $S_0(1)$ at 30.3 kPa and 35 MHz for $S_0(0)$ at 19.3 kPa. The diffusion-model parameters $A$ and $B$, as well as the high-pressure collisional-broadening coefficient $\gamma_2$ are summarized in Table I and compared with prior reports [12,23,24]. We find that $A$ obtained in this study is larger than previously reported values, whereas $B$ remains in good agreement. Our fitted $\gamma_2$ is also larger than that reported in Ref. [30], which we attribute primarily to the substantially lower spectral resolution in earlier spontaneous Raman measurements, where instrumental broadening can bias extracted pressure-broadening coefficients.

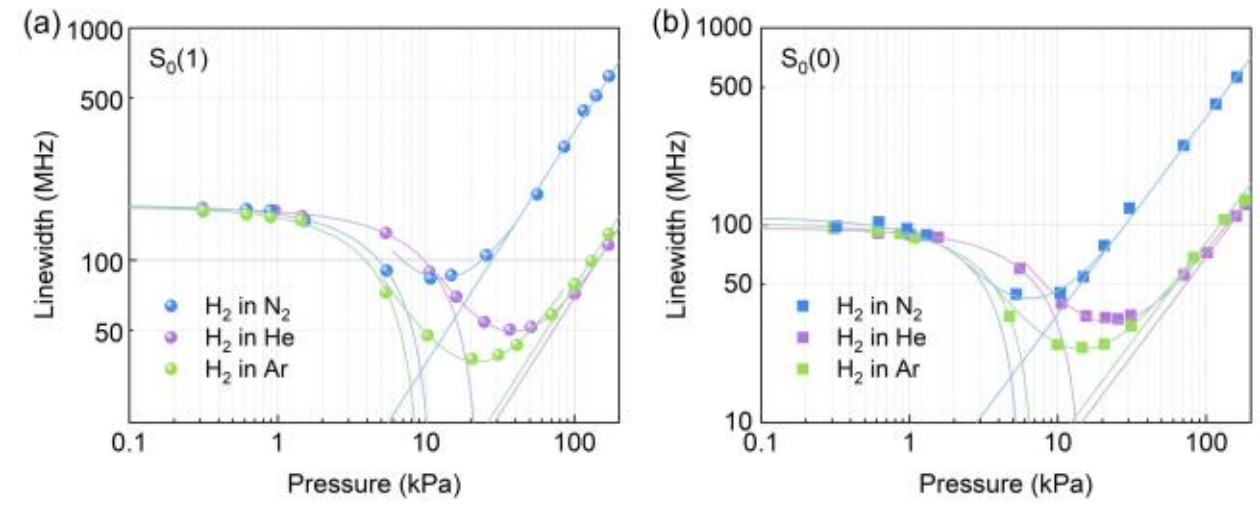

FIG 4. Pressure-dependent linewidths of $H_2$ Raman transitions in different buffer gases ($N_2$, He, or Ar). (a) Variation of the measured FWHM with total gas pressure for $S_0(1)$. (b) Variation of the measured FWHM with total gas pressure for $S_0(0)$. Solid curves are fits using the collisional narrowing, diffusion and collisional broadening models in the specific pressure regimes.

In addition, we evaluated the pressure dependence of the SRS Raman gain from the measured spectra. The resulting Raman gain coefficients are in good agreement with calculations based on the linewidth models (see Supplementary Note 5). In the collisional-narrowed regime, the Stokes gain coefficients for both $S_0(1)$ and $S_0(0)$ increase rapidly with pressure. Beyond the pressure at which the linewidth reaches its minimum, the pressure dependence weakens as collisional broadening gradually dominates the linewidth.

Our spectroscopic system demonstrates a large dynamic range exceeding five orders of magnitude (see Supplementary Note 6). This enables us to further investigate Dicke narrowing in diluted $H_2$ mixtures. Here we studied both $S_0(0)$ and $S_0(1)$ transitions in $H_2$ mixtures buffered by $N_2$, He and Ar. For total gas pressures above 1 kPa, the mole fraction of $H_2$ is fixed at 1%; below 1 kPa, the $H_2$ fraction is increased to maintain an adequate signal-to-noise ratio (SNR). Most spectra measured in this study exhibit SNRs of 100-1000, with the lowest SNR remaining above 40. All gas mixing ratios used in this study are provided in Supplementary Note 7. As shown in Fig. 4, Dicke narrowing is observed for both transitions in all three buffer gases. For each buffer gas, we fit the pressure-dependent linewidths in three pressure intervals using the corresponding regime-specific models (collisional narrowing, diffusion, and collisional broadening); the extracted coefficients are summarized in Table II; representative Raman spectra with their fits are provided in Supplementary Note 8. From the fits, the onset pressures of Dicke narrowing for $S_0(1)$ are 3.17 kPa ($N_2$), 4.47 kPa (He) and 3.36 kPa (Ar), while those for $S_0(0)$ are 1.91 kPa ($N_2$), 2.70 kPa (He) and 2.02 kPa (Ar). All onset pressures fall within our experimental range. To our knowledge, these measurements constitute the first experimental determination of the complete Dicke-narrowing evolution of the $H_2$ $S_0(1)$ and $S_0(0)$ rotational Raman linewidths in the presence of $N_2$, He, and Ar.

TABLE II. Broadening coefficients with fitting uncertainties for $S_0(1)$ and $S_0(0)$ lines of $H_2$ perturbed by $N_2$, He and Ar at 298 K.

| | | $\gamma_1$ (MHz/atm) | A (MHz·atm) | B (MHz/atm) | $\gamma_2$ (MHz/atm) |
|---|---|---|---|---|---|
| $S_0(1)$ | $N_2$ | -1532.20±103.67 | 5.29±0.16 | 330.08±3.16 | 364.03±3.70 |
| | Ar | -1818.54±71.32 | 3.98±0.11 | 83.88±2.27 | 78.01±1.63 |
| | He | -717.32±63.15 | 9.39±0.09 | 64.95±0.76 | 70.05±1.17 |
| $S_0(0)$ | $N_2$ | -1953.96±394.53 | 1.36±0.20 | 333.02±16.59 | 358.29±4.10 |
| | Ar | -1457.64±13.34 | 1.70±0.03 | 81.54±0.46 | 82.33±1.01 |
| | He | -667.14±35.48 | 3.58±0.13 | 72.29±2.06 | 70.03±0.90 |

It is of interest to observe that in $N_2$-buffered mixtures, the Dicke-narrowing signature is rapidly quenched as pressure increases because collisional broadening quickly dominates the linewidth. By contrast, in He, whose molecular mass is closest to that of $H_2$, the fitted coefficients are similar to those measured in pure $H_2$. As a result, the narrowing onset occurs at the highest pressure, and the transition to the collisional-broadening limit also occurs at the highest pressure, among all four gas compositions investigated (including pure $H_2$). Ar exhibits the strongest collisional narrowing, yielding the smallest linewidth for both $S_0(1)$ and $S_0(0)$ at the bottom of the Dicke narrowing region. The narrowing depth, defined as the difference between the Doppler linewidth and the minimum linewidth, as well as the width of the narrowing region and the pressure at which the minimum occurs, vary significantly across buffer gases. These trends highlight that Dicke narrowing in the $H_2$ $S_0(1)$ and $S_0(0)$ rotational Raman transitions is highly sensitive to the collision partner. This strong buffer-gas dependence motivates systematic measurements across additional perturbers and temperatures to benchmark lineshape theories and to refine collisional-transport parameters for $H_2$ Raman spectroscopy.

In conclusion, we demonstrate the first measurement of the complete Dicke-narrowing evolution of the $H_2$ rotational Raman transitions using intracavity dual-resonance stimulated Raman spectroscopy with cavity ringdown readout. This platform uniquely combines kHz-level spectral resolution, high detection sensitivity, and a large dynamic range, which are difficult to achieve simultaneously using conventional Raman techniques. Our experimental data reveal that the continuous linewidth evolution in the intermediate-pressure region is not fully captured by commonly used pressure-dependent models. These results therefore provide stringent experimental benchmarks for developing unified lineshape descriptions that consistently incorporate Dicke narrowing, velocity-changing collisions, and other intermediate-pressure transport and collisional effects.

## ACKNOWLEDGMENTS

This work is supported by Research Impact Fund (R4022-25) of Research Grant Council (RGC), Innovation and Technology Fund (GHP/138/22) of Innovation and Technology Commission, Hong Kong SAR, China; 1+1+1 CUHK–CUHK(SZ)–GDSTC Joint Collaboration Fund Project (2025A0505000079).

Supplementary information

# Intracavity Dual-Resonance Stimulated Raman Spectroscopy with Cavity Ringdown Readout for Resolving Hydrogen Rotational Raman Transitions

Qinxue Nie, Guanda Lyu, Yue Yan, and Wei Ren*

Department of Mechanical and Automation Engineering, The Chinese University of Hong Kong, New Territories, Hong Kong SAR, China

*Corresponding author: *renwei@mae.cuhk.edu.hk*

## Supplementary Note 1: Experimental setup

Fig. S1 illustrates the experimental setup used in this work. Continuous-wave pump and Stokes fields are supplied by two external cavity diode lasers (ECDLs, TOPTICA Photonics) and combined by a wavelength-division multiplexer (WDM1). The copropagating pump and Stokes beams are directed to free space via a fiber collimator (FC1) and mode-matched into a Fabry-Perot optical cavity along a common optical axis. The two beams share an identical optical path and are prepared with parallel linear polarizations. To maintain adequate stimulated Raman scattering (SRS) gain and signal-to-noise ratio (SNR) across the different gas compositions and pressures studied, the incident pump power is adjusted between 0.27 mW and 90 mW, while the incident Stokes power is held at 5 mW. The Stokes wavelength is tuned with a triangular waveform to generate cavity ringdown (CRD) events at a rate of 30 Hz. A small fraction of the cavity-reflected light is directed by the polarizing beamsplitter due to its finite extinction ratio. The reflected pump light is separated by WDM2 and detected by a photodetector (PD1) to generate the Pound-Drever-Hall (PDH) error signal for pump locking to the cavity. The cavity-transmitted Stokes light is separated by WDM3 and captured by PD2. A segment of the PD2 signal is sent to a comparator circuit to trigger the AOM, which rapidly extinguishes the Stokes light and synchronously triggers the data acquisition card to record the CRD signals. For each point in a Raman spectrum, a total of 100 averages are conducted to reduce measurement noise.

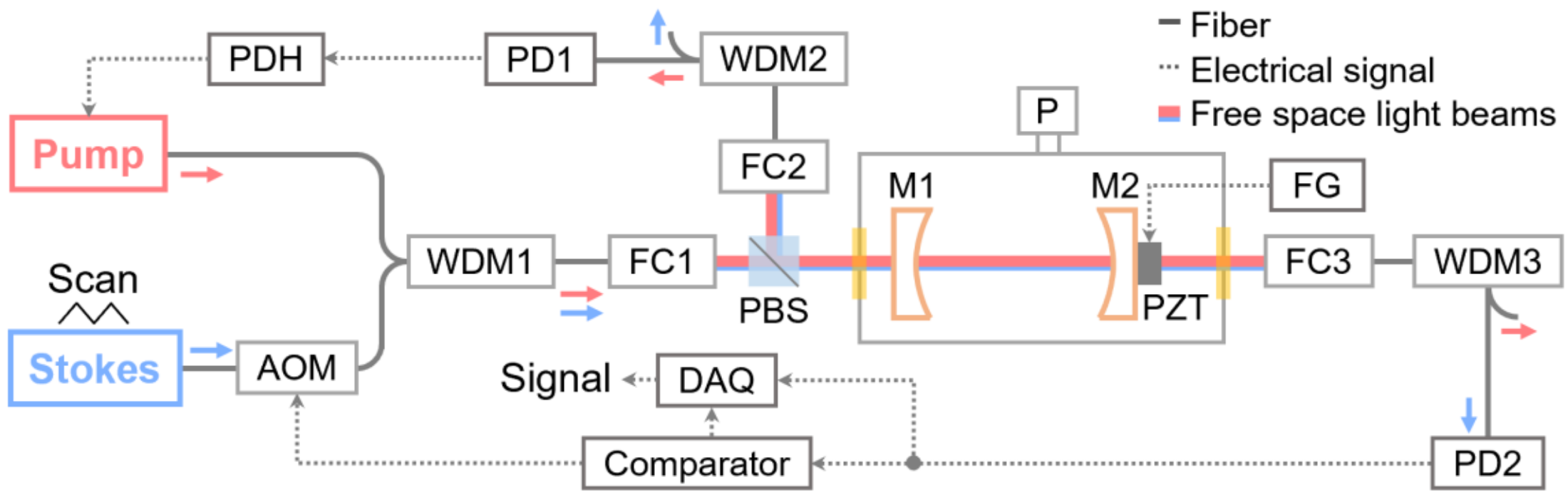

**Fig. S1.** Experimental setup of intracavity double-resonance stimulated Raman spectroscopy with cavity ringdown readout. AOM, acousto-optic modulator; WDM, wavelength-division multiplexer; FC, fiber collimator; PBS, polarizing beamsplitter; M, cavity mirror; PZT, piezoelectric actuator; P, pressure gauge; FG, function generator; PD, photodetector; PDH, Pound-Drever-Hall (PDH) locking module; DAQ, data acquisition system.

## Supplementary Note 2: Characterization of cavity finesse

The reflectivity ($R$) of the cavity mirrors and cavity finesse ($F$) are determined from the empty-cavity ringdown time $\tau_0$:

$$R = 1 - \frac{L}{c\tau_0} \tag{S1}$$

$$F = \frac{\pi\sqrt{R}}{1 - R} \tag{S2}$$

where $L$ is the cavity length and $c$ is the speed of light. The two cavity mirrors are assumed to have identical reflectivities. Representative CRD signals at the pump wavelength and two Stokes wavelengths are plotted in Fig. S2. The corresponding reflectivities are 99.9978% ($F$ = 146,105) at 1530.8 nm, 99.9984% ($F$ = 192,164) at 1681.9 nm and 99.9986% ($F$ = 229,879) at 1618.6 nm.

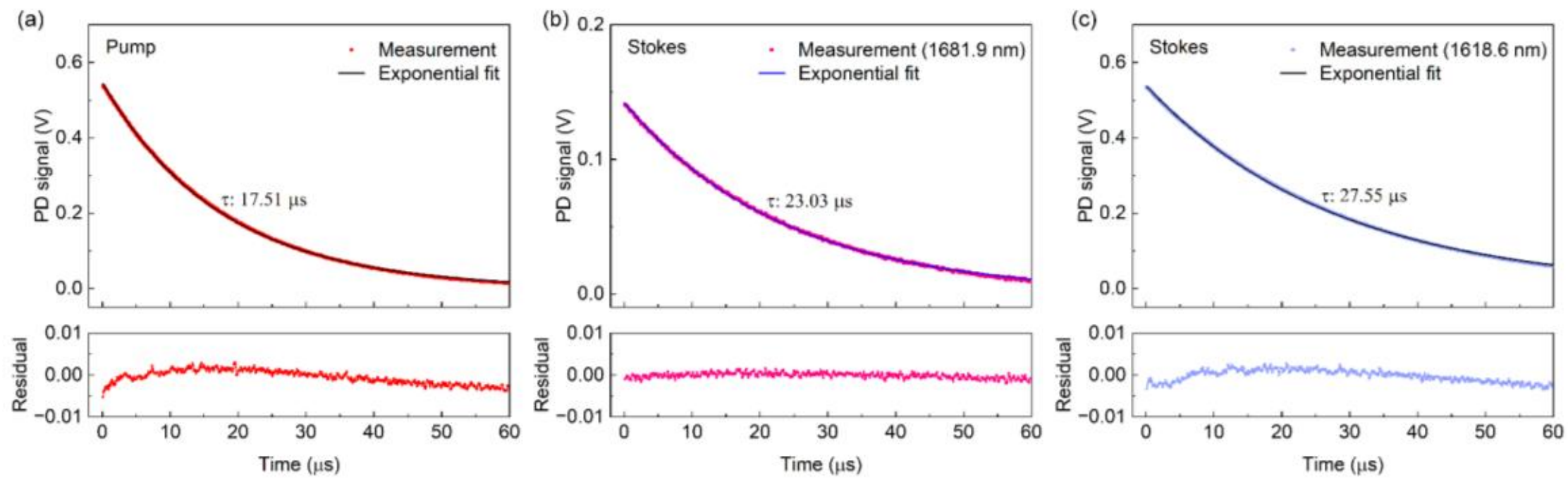

**Fig. S2.** Empty-cavity CRD signals recorded at the wavelengths of (a) 1530.8 nm, (b) 1681.9 nm and (c) 1618.6 nm, respectively. An exponential fit is also provided along with the fitting residual plotted in the bottom panel.

## Supplementary Note 3: Evaluation of Raman spectral resolution

The Raman spectral resolution depends on the minimum tunable frequency difference between the pump and the Stokes lasers. Since both lasers are coupled to and referenced by the same cavity, the pump frequency satisfies $\nu_p = q_p \cdot \nu_{FSR}$ and the Stokes frequency satisfies $\nu_s = q_s \cdot \nu_{FSR}$, where $\nu_{FSR}$ is the free spectral range (FSR), $q_p$ and $q_s$ are the indices of the cavity's longitudinal modes matching the pump and Stokes frequencies, respectively. The frequency difference is therefore:

$$\nu_p - \nu_s = \Delta q \cdot \nu_{FSR} \quad \text{(S3)}$$

where $\Delta q = q_p - q_s$. A small change in $\nu_{FSR}$ yields a change in the Raman detuning of:

$$\Delta(\nu_p - \nu_s) = \Delta q \cdot \Delta\nu_{FSR} \quad \text{(S4)}$$

where $\Delta\nu_{FSR}$ is the tuning resolution of FSR. Because both lasers are locked to cavity longitudinal modes during tuning, the cavity is tuned by following the pump laser. The tuning resolution of FSR is thus determined by the spectral resolution of the frequency-locked pump laser, which in turn is bounded by the cavity mode linewidth $\Gamma_c$ at the pump wavelength ($\Delta\nu_{FSR} = \Gamma_c/q_p$). Substituting it into Eqn. S4 yields:

$$\Delta(\nu_p - \nu_s) = \Delta q \cdot \Gamma_c/q_p \quad \text{(S5)}$$

indicating that the Raman spectral resolution is reduced relative to the pump optical frequency scale by a factor $\Delta q/q_p$. The mode parameters used for measuring $H_2$ $S_0(1)$ and $S_0(0)$ rotational Raman transitions are listed in Table S1. The resulting minimum attainable Raman spectral resolutions are 0.49 kHz for $S_0(1)$ and 0.82 kHz for $S_0(0)$ according to Eqn. (S5).

**Table S1.** Cavity mode parameters used for measuring $H_2$ $S_0(1)$ and $S_0(0)$ transitions.

| Transition | $q_p$ | $q_s$ | $\Delta q$ | $\Gamma_c$ at $\nu_p$ | $\Delta(\nu_p - \nu_s)$ |
|---|---|---|---|---|---|
| $S_0(1)$ | 147572 | 134311 | 13261 | 9.1 kHz | 0.49 kHz |
| $S_0(0)$ | 147572 | 139567 | 8005 | 9.1 kHz | 0.82 kHz |

## Supplementary Note 4: Representative Raman spectra of pure $H_2$ at different pressures

**Error! Reference source not found.** and Fig. S4 depict representative spectra of $H_2$ $S_0(1)$ and $S_0(0)$ across the pressure regimes investigated. At low pressure (Doppler-broadened regime), the spectra are well fitted by Gaussian profiles, as shown in Fig. S3(a) and Fig. S4(a). At intermediate pressures where collisional broadening becomes appreciable while Doppler broadening remains significant, the spectra are described by Voigt profiles, as shown in Fig. S3(b) and Fig. S4(b). At higher pressures, up to the collisional-broadening regime, Lorentzian profiles adequately describe the measured lineshapes, as shown in Fig. S3(c,d,e,f) and Fig. S4(c,d,e,f).

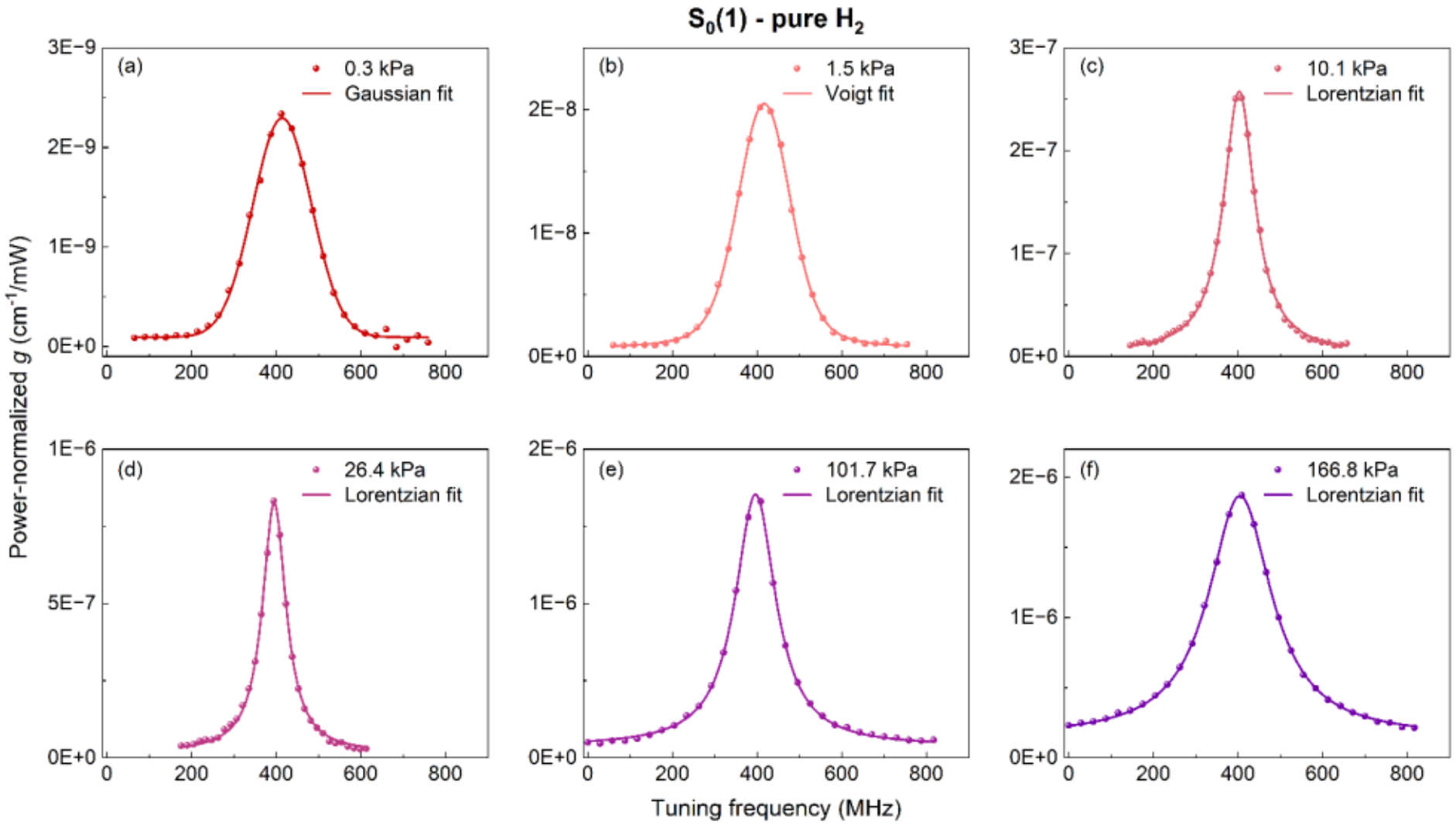

**Fig. S3.** Representative Raman spectra of $H_2$ $S_0(1)$ transition measured at different pressures with the corresponding fitting profiles.

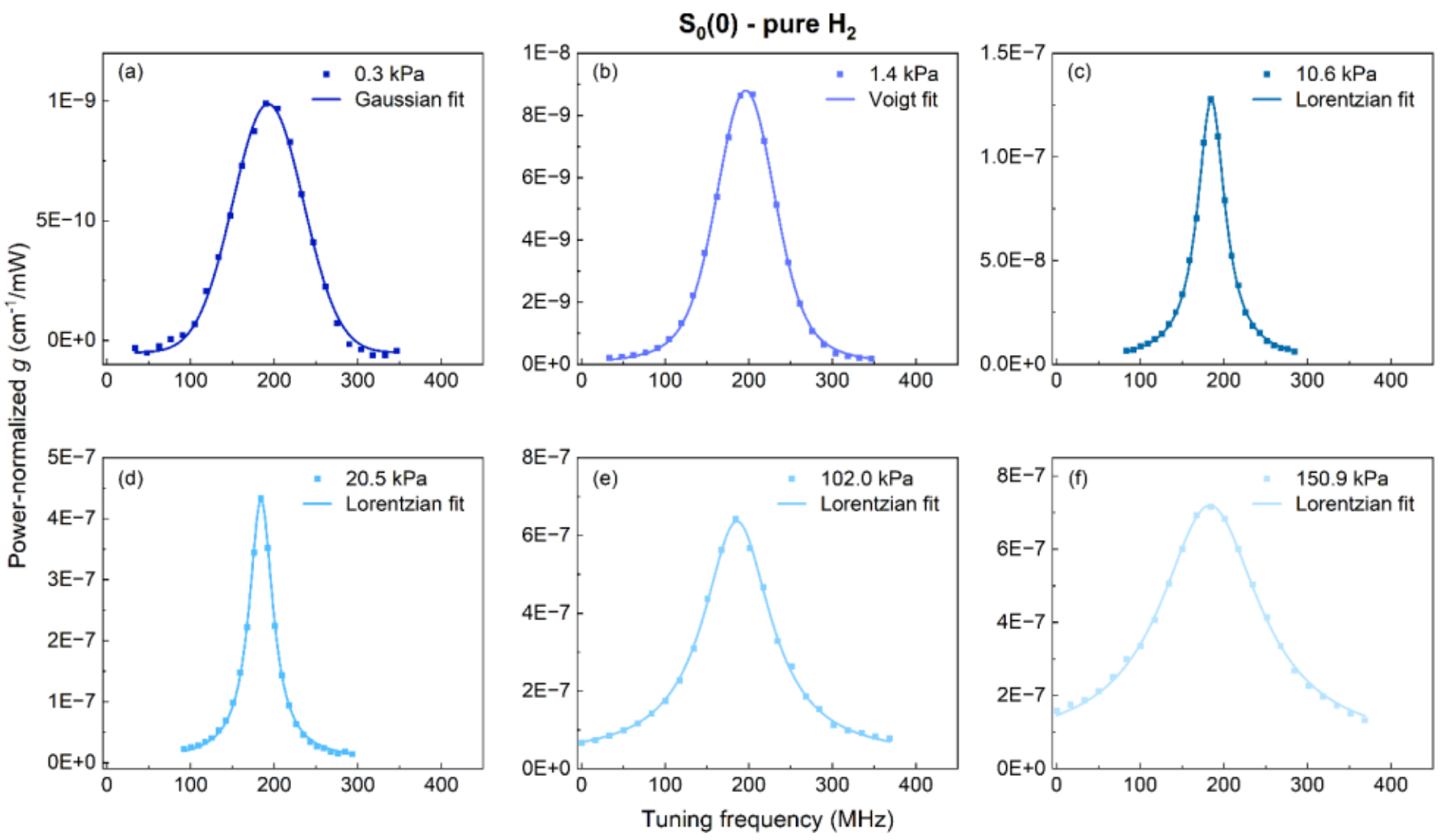

**Fig. S4.** Representative Raman spectra of $H_2$ $S_0(0)$ transition measured at different pressures with the corresponding fitting profiles.

## Supplementary Note 5: Raman gain coefficient determination in the Dicke narrowing regime

The measurements shown in Fig. 3 of the main text allow determination of the pressure dependence of the Raman gain coefficient. We ensure SNR of the CRD signals over the full pressure range by adjusting the pump laser power, because the Stokes gain coefficient scales linearly with the pump power. Particularly, the pump power is adjusted from 0.27 mW at the highest pressure to 90 mW at low pressures. As a result, the measured Stokes gain coefficients are normalized to the incident pump power to obtain the Raman gain coefficient. The resulting pressure-dependent gain coefficients for $S_0(1)$ and $S_0(0)$ are depicted in Fig. S5(a) and (b), respectively. For a gas-phase Raman transition $S_0(J)$, the gain coefficient $g_R$ is given by [1]:

$$g_R(S_0(J)) = \frac{32\pi^3 \nu_s}{c_0^2} \frac{\Delta N_{fi}}{h \Delta \nu_R} \frac{1}{5} \frac{(J+1)(J+2)}{(2J+1)(2J+3)} (\gamma_{00}^2) \tag{S6}$$

where $J$ is the rotational quantum number, $c_0$ is the speed of light, $h$ is Planck's constant, $\Delta\nu_R$ is the linewidth of the Raman transition, and $\gamma_{00}$ is the molecular polarizability. The population difference $\Delta N_{fi}$ between the initial state ($i$) and final state ($f$) is proportional to gas pressure. Using the measured pressure-dependent linewidths (Fig. 3), the corresponding pressure dependence of $g_R$ can be calculated and are illustrated in Fig. S5. The measured Raman gain coefficients show good agreement with the calculations based on the three linewidth models.

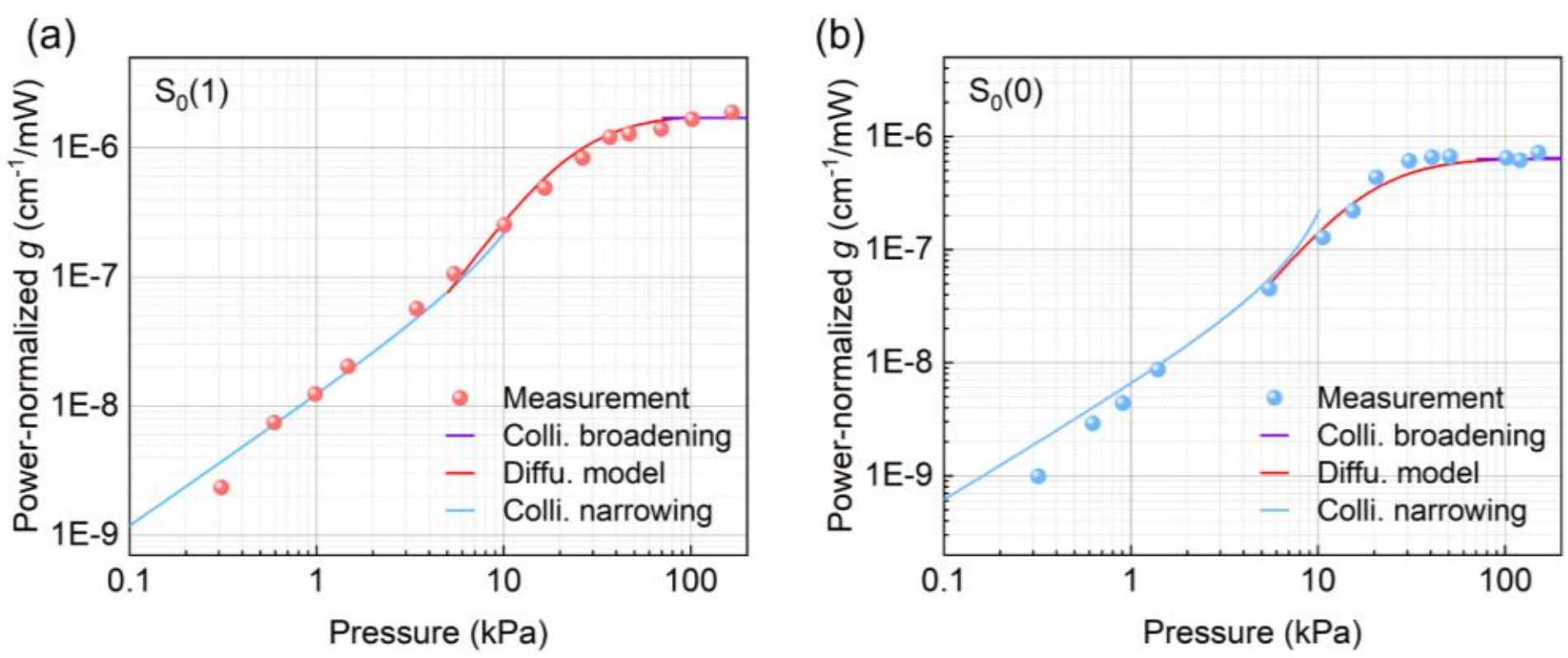

**Fig. S5.** Pressure dependence of the SRS gain coefficient for (a) $S_0(1)$ and (b) $S_0(0)$ transitions, together with the corresponding calculation results.

## Supplementary Note 6: Characterization of dynamic range and detection sensitivity

To characterize the measurement dynamic range, the SRS gain is normalized by the cavity incident pump power, exploiting the linear scaling of the SRS signal with pump power. The variation of power-normalized SRS gain coefficient with gas pressure can be found in Supplementary Note 5. Fig. S6 demonstrates the evaluation of the minimum detection limit using Allan deviation analysis. For this analysis, pure $N_2$ is continuously monitored while the pump laser remains locked to the optical cavity at an incident power of 90 mW. The Stokes laser is repeatedly scanned across the $S_0(1)$ and $S_0(0)$ transitions, and more than 2600 Stokes CRD events are recorded at each Stokes wavelength. Based on the upper limits of the SRS gain coefficients from Supplementary Note 5, the dynamic range for $S_0(1)$ spans $2 \times 10^{-6}$ to $2 \times 10^{-12}$ $cm^{-1}/mW$ (six orders of magnitude) and $7 \times 10^{-7}$ to $1.5 \times 10^{-12}$ $cm^{-1}/mW$ for $S_0(0)$ (over five orders of magnitude).

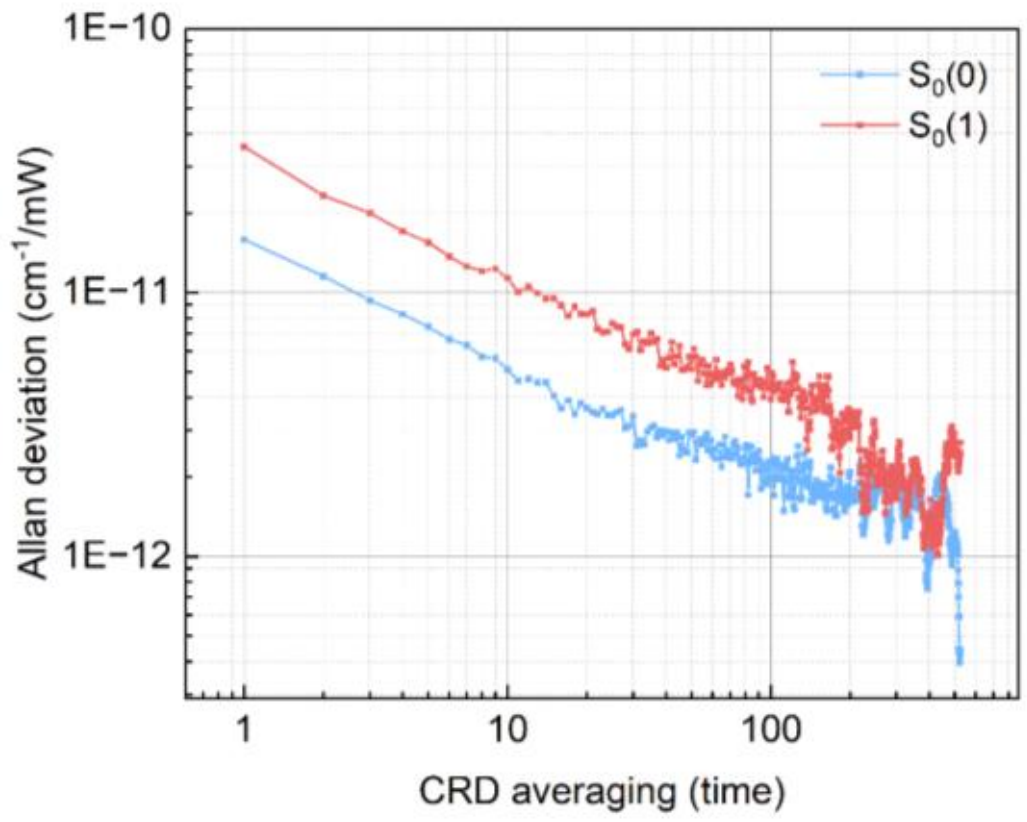

**Fig. S6.** Evaluation of the minimum detection limit by Allan deviation analysis.

## Supplementary Note 7: Composition of $H_2$–buffer gas mixtures

$H_2$–buffer gas mixtures are prepared by diluting high-purity $H_2$ (99.999%, Linde) with high-purity $N_2$ (99.995%, Linde), He (99.995%, Linde), or Ar (99.95%, Linde). Gas mixtures with different compositions, which are listed in Table S2 and Table S3, are used to maintain sufficient measurement SNR. The linewidths of the $H_2$-buffer gas are extracted from the measurements by subtracting the $H_2$-$H_2$ linewidths using the following equation:

$$\Gamma_{H_2-i} = \frac{\Gamma_{meas} - \Gamma_{H_2-H_2} * X}{1 - X} \tag{S7}$$

where $X$ is the $H_2$ mole fraction in the gas mixture, $\Gamma_{H_2-i}$ is the linewidth of $H_2$ in different buffer gases ($i$ denotes the buffer gas such as $N_2$, He, and Ar), $\Gamma_{meas}$ is the directly measured linewidth of the mixture, and $\Gamma_{H_2-H_2}$ is the linewidth of pure $H_2$ (from Fig. 3 in the main text).

**Table S2.** Gas mixtures used for $H_2$ $S_0(1)$ transition measurements.

| Pressure (kPa) | $H_2$ fraction in $N_2$ (%) | Pressure (kPa) | $H_2$ fraction in He (%) | Pressure (kPa) | $H_2$ fraction in Ar (%) |
|---|---|---|---|---|---|
| 171.84 | 1 | 170.04 | 1 | 170.66 | 1 |
| 140.55 | 1 | 100.81 | 1 | 130.7 | 1 |
| 115.61 | 1 | 70.36 | 1 | 100.47 | 1 |
| 85.61 | 1 | 51.11 | 1 | 70.65 | 1 |
| 56.21 | 1 | 36.72 | 1 | 41.11 | 1 |
| 25.53 | 1 | 24.73 | 1 | 30.79 | 1 |
| 14.85 | 1 | 16.00 | 1 | 20.44 | 1 |
| 10.75 | 1 | 10.57 | 1 | 10.24 | 1 |
| 5.425 | 1 | 5.37 | 1 | 5.38 | 1 |
| 1.53 | 5 | 1.463 | 5 | 1.44 | 5 |
| 0.891 | 10 | 0.959 | 10 | 0.896 | 10 |
| 0.617 | 20 | 0.621 | 20 | 0.617 | 20 |
| 0.315 | 20 | 0.312 | 20 | 0.315 | 20 |

**Table S3.** Gas mixtures used for $H_2$ $S_0(0)$ transition measurements.

| Pressure (kPa) | $H_2$ fraction in $N_2$ (%) | Pressure (kPa) | $H_2$ fraction in He (%) | Pressure (kPa) | $H_2$ fraction in Ar (%) |
|---|---|---|---|---|---|
| 161.2 | 1 | 187.43 | 1 | 182.22 | 1 |
| 116.26 | 1 | 159.67 | 1 | 132.39 | 1 |
| 70.66 | 1 | 101.59 | 1 | 82.25 | 1 |
| 30.4 | 1 | 70.91 | 1 | 31.16 | 1 |
| 20.553 | 1 | 31.04 | 1 | 20.56 | 1 |
| 14.89 | 1 | 25.42 | 1 | 14.59 | 1 |
| 10.4 | 1 | 20.90 | 1 | 9.959 | 1 |
| 5.25 | 1 | 15.53 | 1 | 4.718 | 1 |
| 1.305 | 10 | 10.57 | 1 | 1.085 | 10 |
| 0.965 | 10 | 5.55 | 1 | 0.864 | 10 |
| 0.621 | 20 | 1.558 | 10 | 0.607 | 50 |
| 0.322 | 50 | 1.022 | 20 | 0.306 | 50 |
| | | 0.617 | 50 | | |
| | | 0.310 | 50 | | |

## Supplementary Note 8: Representative Raman spectra of $H_2$ in buffer gas mixtures

Representative spectra of the $H_2$ $S_0(1)$ transition measured in $H_2$–$N_2$, $H_2$–He, and $H_2$–Ar mixtures at different pressures are demonstrated in Fig. S7, Fig. S8 and Fig. S9, respectively. Corresponding spectra for the $S_0(0)$ transition are illustrated in Fig. S10, Fig. S11 and Fig. S12. As described in Supplementary Note 7, the $H_2$ concentration is increased at low pressures to maintain a sufficient SNR for spectral fitting. Therefore, all displayed Raman gains are normalized to the equivalent value for 1% $H_2$. The $H_2$ fraction used in each measurement is indicated in parentheses in the figure legend. The narrowest linewidth of $S_0(1)$ in the $H_2$–$N_2$ mixture occurs at 10.8 kPa, whereas that of $S_0(0)$ occurs at 5.3 kPa. In the $H_2$-He mixture, the minima occur at 36.7 kPa for $S_0(1)$ and 25.4 kPa for $S_0(0)$. In the $H_2$–Ar mixture, the minima occur at 20.4 kPa for $S_0(1)$ and 14.6 kPa for $S_0(0)$.

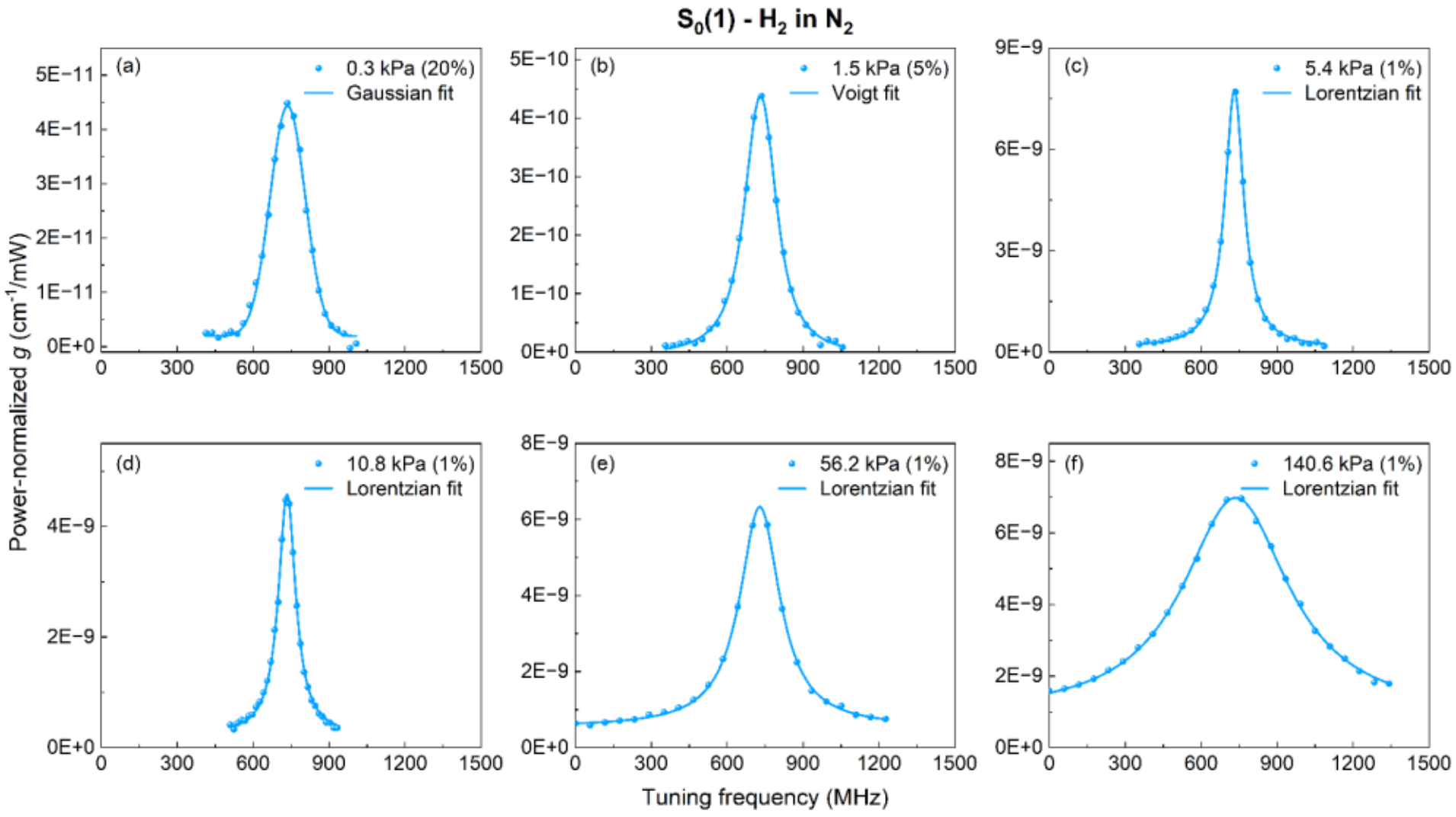

**Fig. S7.** Representative Raman spectra of the $H_2$ $S_0(1)$ transition measured in $H_2$–$N_2$ mixtures across different pressure regimes with their fitting profiles.

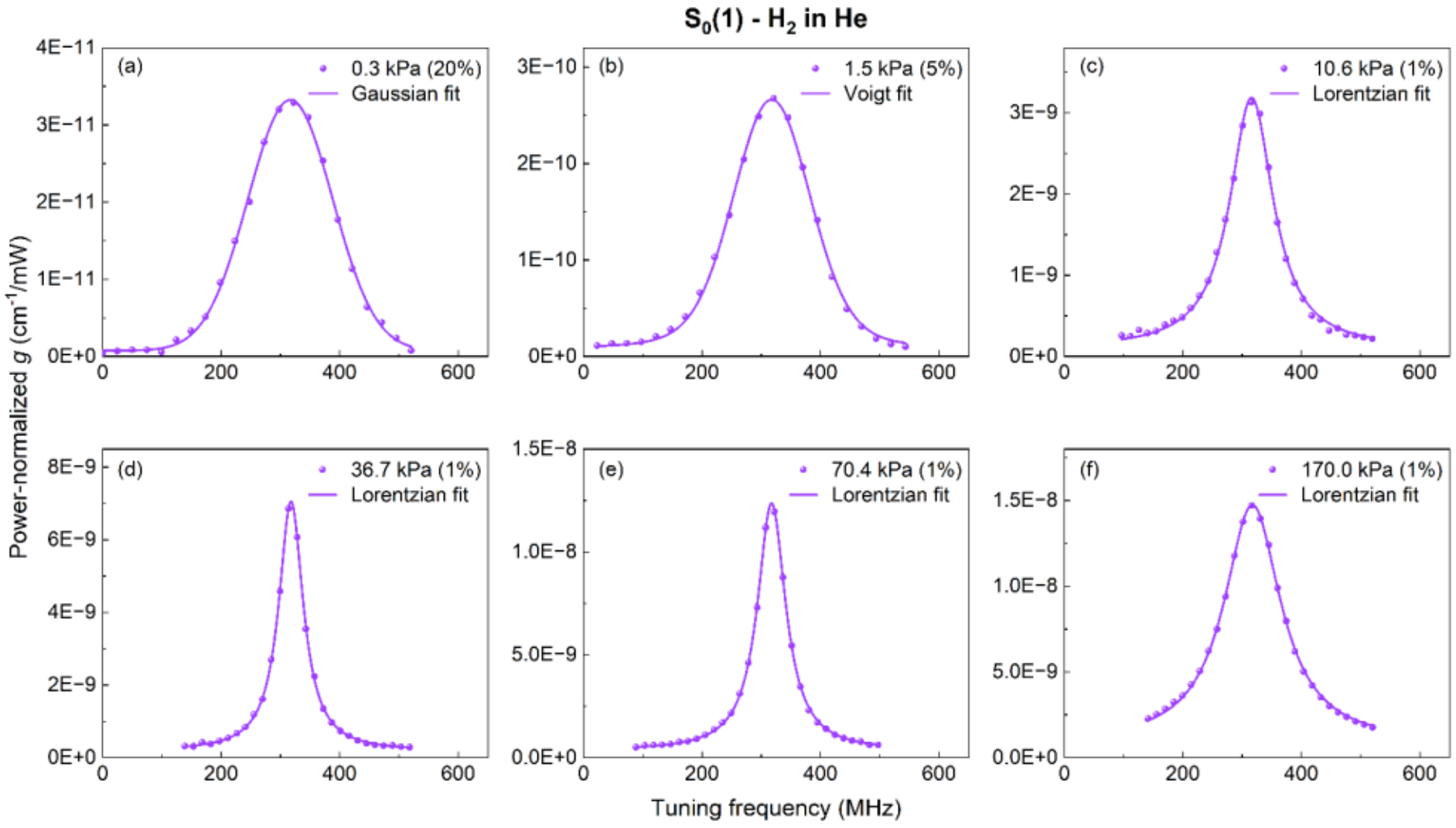

**Fig. S8.** Representative Raman spectra of the $H_2$ $S_0(1)$ transition measured in $H_2$–He mixtures across different pressure regimes with their fitting profiles.

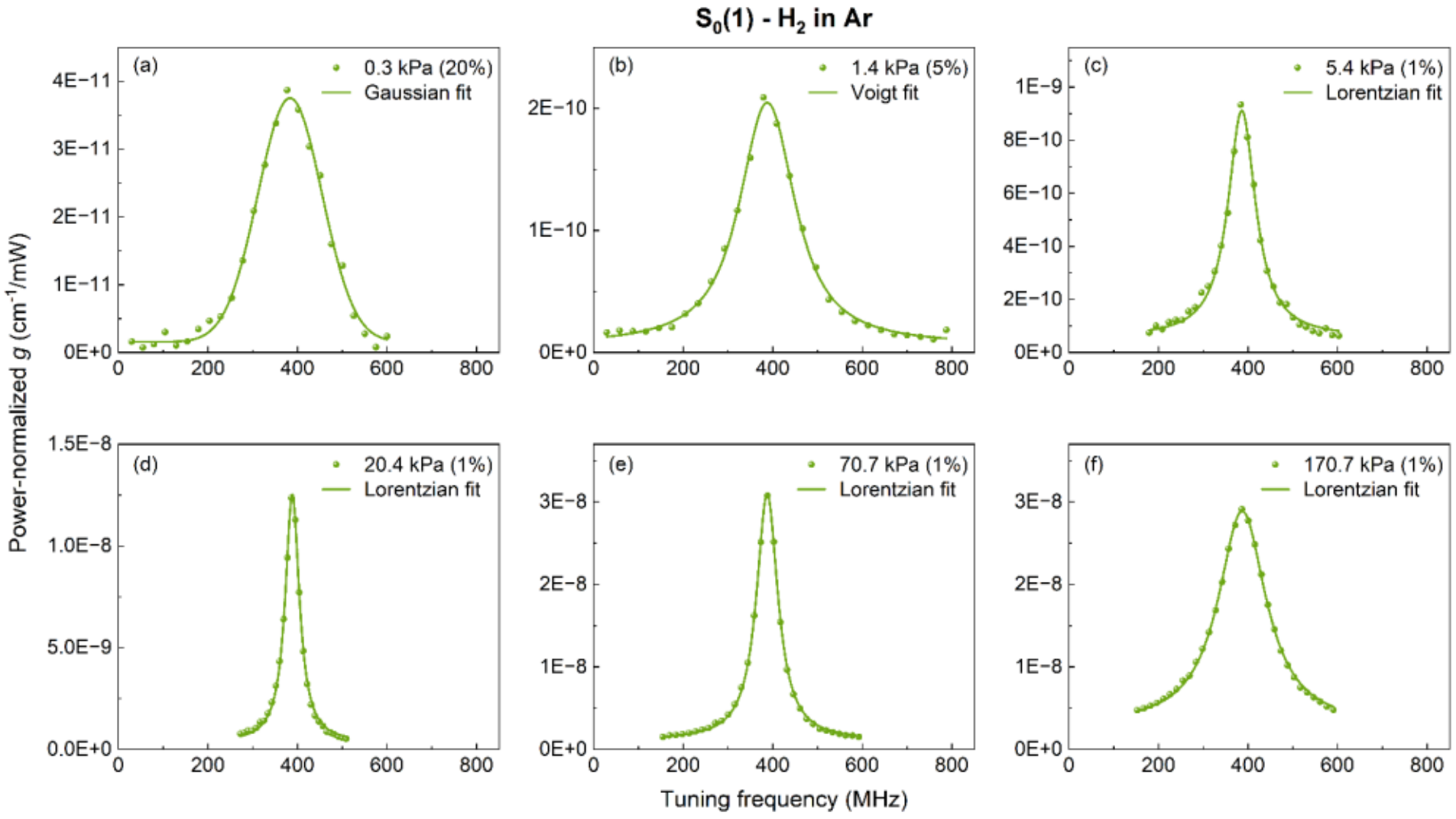

**Fig. S9.** Representative Raman spectra of the $H_2$ $S_0(1)$ transition measured in $H_2$–Ar mixtures across different pressure regimes with their fitting profiles.

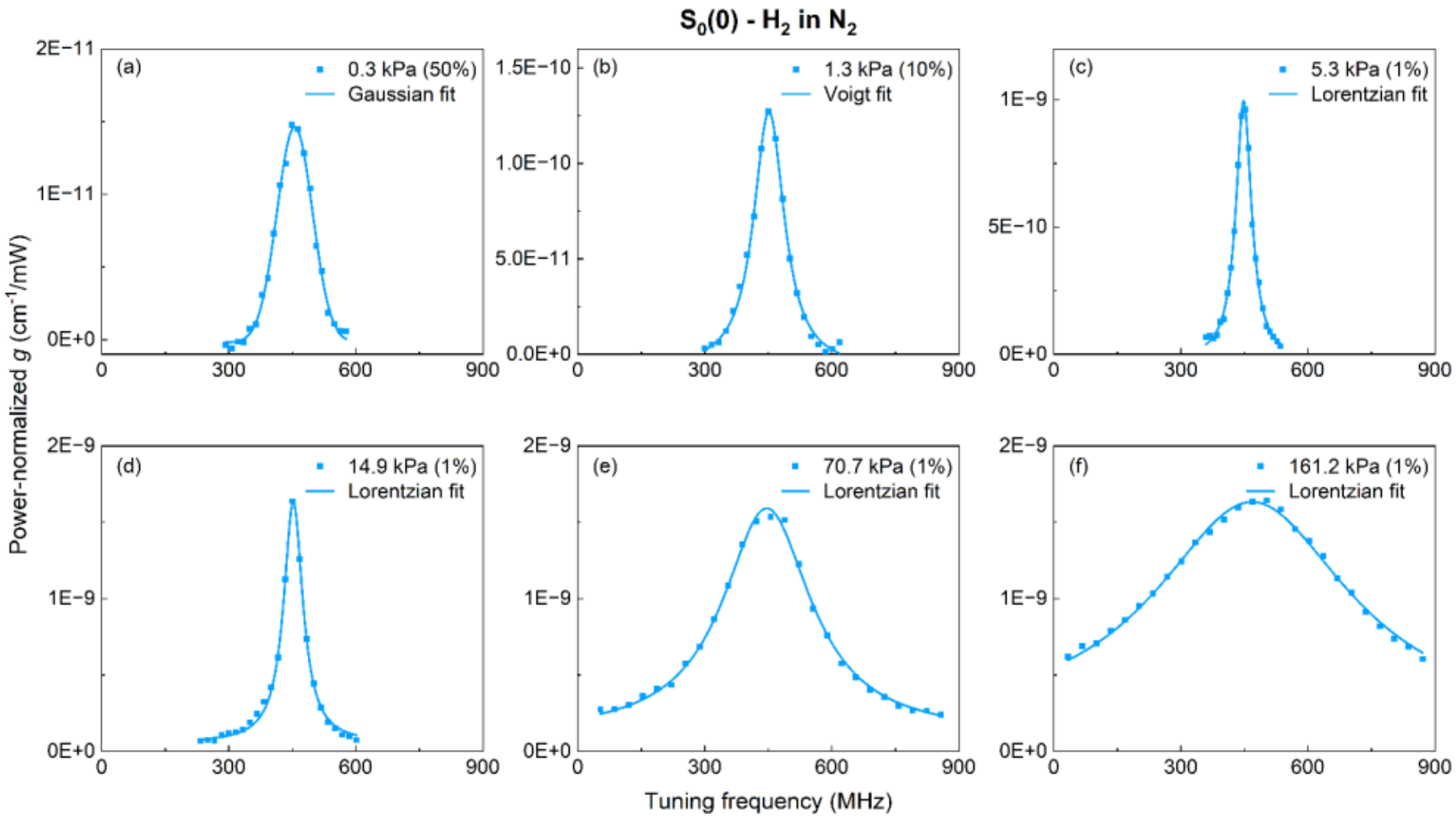

**Fig. S10.** Representative Raman spectra of the $H_2$ $S_0(0)$ transition measured in $H_2$–$N_2$ mixtures across different pressure regimes with their fitting profiles.

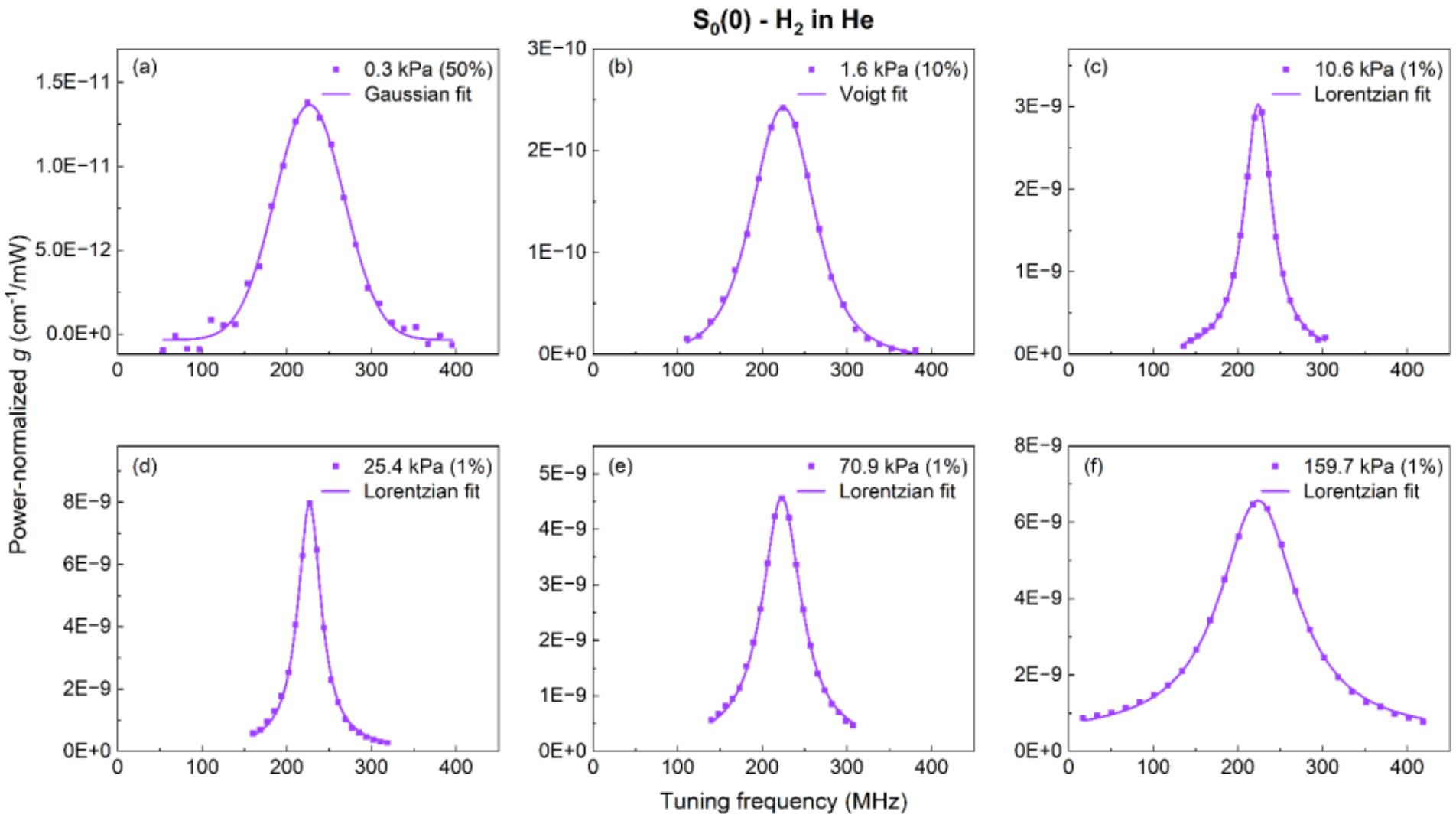

**Fig. S11.** Representative Raman spectra of the $H_2$ $S_0(0)$ transition measured in $H_2$–He mixtures across different pressure regimes with their fitting profiles.

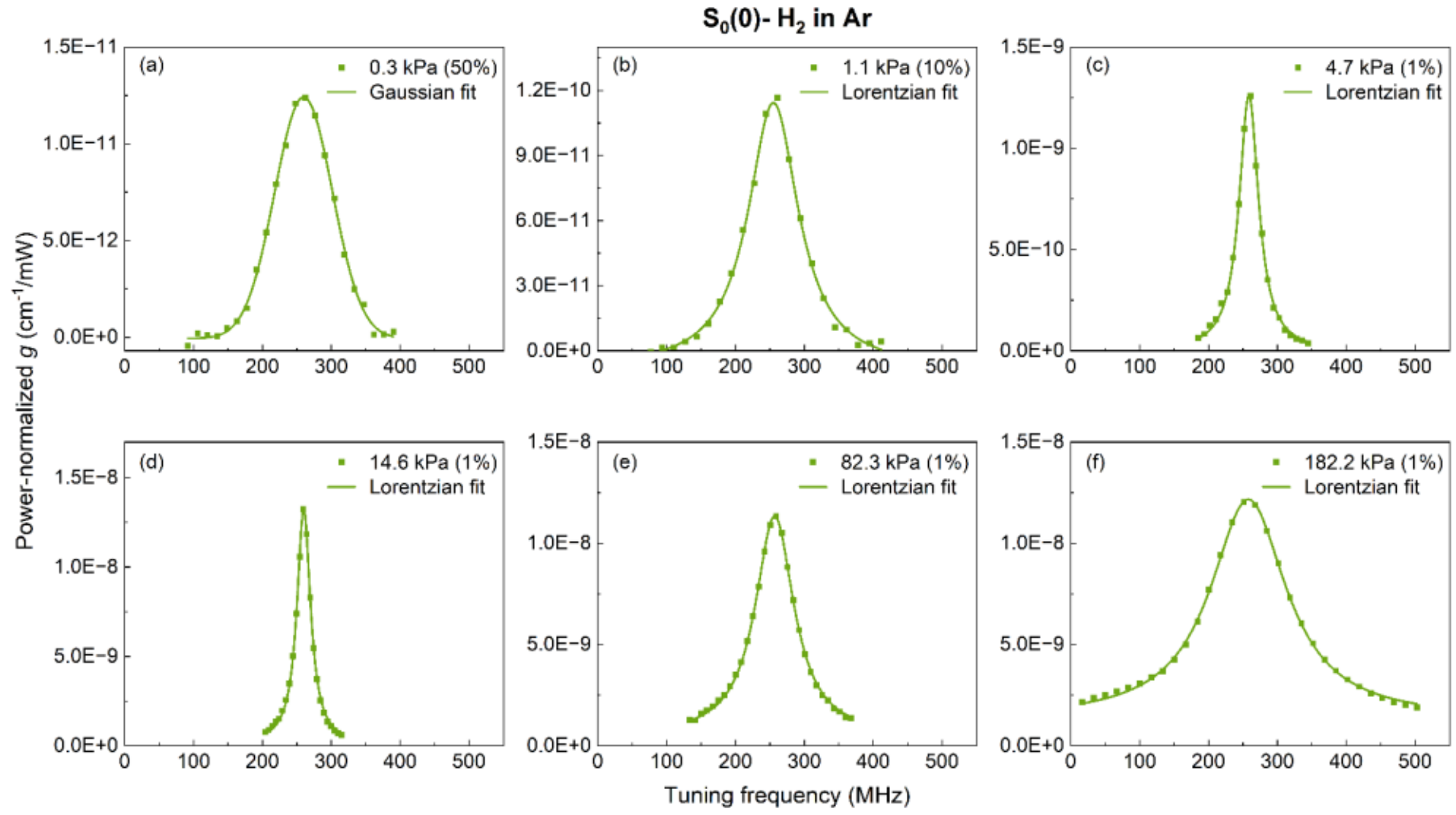

**Fig. S12.** Representative Raman spectra of the $H_2$ $S_0(0)$ transition measured in $H_2$–Ar mixtures across different pressure regimes with their fitting profiles.